\documentclass[twocolumn,superscriptaddress,amssymb,amsmath,floatfix,citeautoscript]{revtex4-1}
\usepackage[pdftex]{graphicx}
\usepackage{chngcntr}
\usepackage{amsmath}
\usepackage{color}
\usepackage{natbib}
\usepackage{mathtools}
\usepackage{physics}
\usepackage[title]{appendix}
\allowdisplaybreaks

\begin{document}

\title{Dynamics of nonequilibrium magnons in gapped Heisenberg antiferromagnets}


\author{Chengyun Hua}
\email{To whom correspondence should be addressed. E-mail: huac@ornl.gov}
\author{Lucas Lindsay}
\author{Yuya Shinohara}
\affiliation{Materials Science and Technology Division, Oak Ridge National Laboratory, Oak Ridge, TN 37831, USA}
\author{David Alan Tennant}
\affiliation{Department of Physics and Astronomy, University of Tennessee, Knoxville, TN 37996, USA }
\affiliation{Department of Materials Science and Engineering, University of Tennessee, Knoxville TN, 37996, USA}
\affiliation{Shull Wollan Center, Oak Ridge National Laboratory, Oak Ridge TN, 37831, USA }

\date{\today}

\widetext

Notice:  This manuscript has been authored by UT-Battelle, LLC, under contract DE-AC05-00OR22725 with the US Department of Energy (DOE). The US government retains and the publisher, by accepting the article for publication, acknowledges that the US government retains a nonexclusive, paid-up, irrevocable, worldwide license to publish or reproduce the published form of this manuscript, or allow others to do so, for US government purposes. DOE will provide public access to these results of federally sponsored research in accordance with the DOE Public Access Plan (http://energy.gov/downloads/doe-public-access-plan).

\thispagestyle{empty}

\newpage

\clearpage
\pagenumbering{arabic}

\begin{abstract}

Nonequilibrium dynamics in spin systems is a topic currently under intense investigation as it provides fundamental insights into thermalization, universality, and exotic transport phenomena. While most of the studies have been focused on ideal closed quantum many-body systems such as ultracold atomic quantum gases and one-dimensional spin chains, driven-dissipative Bose gases in steady states away from equilibrium in classical systems also lead to intriguing nonequilibrium physics. In this work, we theoretically investigate out-of-equilibrium dynamics of magnons in a gapped Heisenberg quantum antiferromagnet based on Boltzmann transport theory. We show that, by treating scattering terms beyond the relaxation time approximation in the Boltzmann transport equation, energy and particle number conservation mandate that nonequilibrium magnons cannot relax to equilibrium, but decay to other nonequilibrium stationary states, partially containing information about the initial states. The only decay channel for these stationary states back to equilibrium is through the non-conserving interactions (\emph{i.e.}, changing particle number and/or energy within the magnon system) such as boundary or magnon-phonon scattering. At low temperatures, these non-conserving interactions are much slower processes than intrinsic magnon-magnon interaction in a gapped spin system. Using magnon-phonon interaction as a quintessential type of non-conserving interaction, we then propose that nonequilibrium steady states of magnons can be maintained and tailored using periodic driving at frequencies faster than relaxation due to phonon interactions. These findings reveal a class of classical material systems that are suitable platforms to study nonequilibrium statistical physics and macroscopic phenomena such as classical Bose-Einstein condensation of quasiparticles and magnon supercurrents that are relevant for spintronic applications. 

\end{abstract}


\maketitle

\section{Introduction}

Nonequilibrium dynamics in spin systems is important as it opens the door to exciting new fundamental scientific questions about thermalization, universality, and dynamical phase transitions beyond traditional condensed matter paradigms \cite{vasseur_nonequilibrium_2016}. Technologically, it might hold the key to achieving coherent spin transport as it directly relates to dissipationless information transport for low-power and low-loss spintronic devices \cite{hortensius_coherent_2021}. Strongly motivated by the rapid progress in achieving closed quantum system such as ultra-cold atomic, molecular, and trapped ion systems, recent theoretical studies of nonequilibrium spin dynamics have been focused on closed many-body quantum systems \cite{myers_transport_2020, bulchandani_solvable_2017,wei_quantum_2022,doyon_thermalization_2017,rigol_relaxation_2007, kinoshita_quantum_2006}, such as one-dimensional (1D) spin chains \cite{scheie_detection_2021,bertini_transport_2016, bertini_low-temperature_2018,essler_quench_2016}. In 1D quantum systems, the ubiquitous presence of integrability---a property of dynamical systems with an infinite set of conserved commuting quantities---leads to intriguing physics such as nonequilibrium flows \cite{bertini_transport_2016, PhysRevX.6.041065,PhysRevB.96.115124,PhysRevB.98.075421}, finite Drude weight \cite{zotos_finite_1999,PhysRevB.96.081118, 10.21468/SciPostPhys.3.6.039, PhysRevLett.119.020602, PhysRevB.97.045407}, relaxation to nonequilibrium steady states (NESS) \cite{ikeda_general_2020, oka_floquet_2019}, all constrained by the conservation laws afforded by integrability. 

Spin dynamics in higher dimensional quantum magnets, \emph{e.g.}, two-dimensional (2D) or three-dimensional (3D) Heisenberg antiferromagnets have also been extensively investigated, both theoretically and experimentally \cite{cottam_high_1970, balcar_spin-wave_1971, k_p_bohnen_high-frequency_1974, tyc_dynamic_1989, ty_damping_1990, kopietz_magnon_1990}. These systems are well described by a classical spin model where the spin components have a range of possible values depending on their particular angular orientation. In such systems, the collective excitations of spins called magnons act as a dilute Bose gas. Nonequilibrium dynamics of such a Bose gas may also lead to interesting physics. Specifically, in the quantum degenerate regime, an ideal Bose gas driven to a steady state away from equilibrium could form Bose-Einstein condensation\cite{demokritov_boseeinstein_2006, bozhko_supercurrent_2016, borisenko_direct_2020}. This coherent quantum state of magnons holds the key to achieving dissipationless and possibly entangled information transport, realizing the true potential of spintronics for low power and low loss electronic devices. In this context one might ask simple questions: what are the dynamical properties of nonequilibrium magnons in higher dimensional quantum magnets? In particular, under what conditions could they reach a NESS?

A conventional understanding of magnon dynamics is that the dynamics is dominated by collisions among magnons, where the transition rate is given by Fermi's golden rule \cite{van_hove_quantum-mechanical_1954, van_hove_approach_1957}. Therefore, nonequilibrium magnons, whose statistics are not described by the Bose-Einstein distribution, are believed to relax back to equilibrium states due to these collision processes. Theoretical investigations using the Boltzmann description of transport under the relaxation time approximation (RTA) supports such a claim \cite{cornelissen_magnon_2016,zhang_spin_2012, zhang_magnon_2012, althammer_allelectrical_2021,liu_collective_2019}. However, the validity of the RTA should be questioned when multiple macroscopic conservation laws are imposed, thus limiting interaction pathways. 

For higher dimensional Heisenberg antiferromagnets at low temperatures, the symmetry of spin operators restricts intrinsic magnon-magnon interactions to elastic pairwise collisions of magnons, meaning microscopic interactions conserve crystal momentum, energy, and particle number \cite{harris_dynamics_1971} and macroscopically total energy and particle number are conserved.Non-conserving interactions (\emph{i.e.}, changing particle number and/or energy within the magnon system) such as magnon-phonon scatterings are very weak in a gapped system at low temperatures, as the overlap of scattering phase space between phonon and magnon states is significantly reduced compared to gapless systems. Boundary scatterings are also very weak in a gapped system since the occupied states at low temperatures are  primarily near the zone center and their group velocities are close to zero. Therefore, these non-conserving interactions occur at a time scale much slower than intrinsic magnon-magnon interactions. This prediction has been confirmed by high-precision neutron spin echo measurements of magnon linewidths of Rb$_2$MnF$_4$ and MnF$_2$, 2D and 3D Heisenberg antiferromagnets, respectively \cite{huberman_study_2008,bayrakci_lifetimes_2013}.

In this work, we first show that, by treating scattering terms beyond the RTA in the Boltzmann transport equation, the conservation of total energy and particle number mandate that nonequilibrium magnons cannot relax to equilibrium but decay to nonequilibrium stationary states whose conditions are partially determined by the initial states. The only instrinsic decay channel for these stationary states back to equilibrium is through the interaction with phonons, which is a much slower process. We then propose that the NESS of magnons can be observed experimentally if such a system is driven by pumping nonequilibrium magnons periodically at a rate faster than the phonon-magnon decay times.

\section{Magnon Boltzmann Transport Equation} 

Out-of-equilibrium magnon transport in Heisenberg quantum magnets without an external magnetic field can be generally described by the Boltzmann transport equation\cite{liu_collective_2019}
\begin{equation}\label{eq:mBTE}
\frac{\partial f_{k}}{\partial t}+\mathbf{v}_k\cdot \nabla_\mathbf{x} f_{k} = -\frac{\partial f_{k}}{\partial t}\biggm|_{\mathrm{scattering}},
\end{equation}
which describes the dynamics of the out-of-equilibrium occupation function $f_{k}$ at position $\mathbf{x}$ and time $t$, for all possible magnon states $k$ ($k \equiv (\mathbf{q},s)$, where $\mathbf{q}$ is the magnon wavevector and $s$ is the magnon polarization). The second term on the left hand side of Eq.~(\ref{eq:mBTE}) describes advection processes of magnons, where $\mathbf{v_{k}}$ is the group velocity derived from the slope of the magnon dispersion. The right hand side term describes variations due to magnon scatterings. 

The spin-wave dispersion and damping in quantum magnets are derived from the second order and higher order expansions of a spin operator Hamiltonian, respectively. At low temperatures, the spin order in a Heisenberg model is collinear and magnon scatterings in a Heisenberg antiferromagnet are restricted to two-in/two-out processes ($k+p \rightarrow s+r$) due to symmetry of the spin operators. Three-magnon scatterings only become important when the non-collinear terms are non-negligible at elevated temperatures, \emph{i.e.} close to the N$\acute{\text{e}}$el temperature\cite{huberman_study_2008,harris_dynamics_1971}, or for frustration. Derived from quantum perturbation theory, the scattering term in Eq.~\ref{eq:mBTE} for two-in/two-out processes is given in Eq.~(\ref{eq:scattering}).  

To solve Eq.~(\ref{eq:mBTE}), we first define the out-of-equilibrium occupation function as 
\begin{equation}
f_{k}(t,\mathbf{x}) = f^{BE}_k(T_0) +\Delta f_k(t,\mathbf{x}),
\end{equation}
where $\Delta f_k(t,\mathbf{x})$ is the deviational distribution from the global equilibrium Bose-Einstein distribution, $f^{BE}_{k}(T_0) = (\text{exp}(\hbar\omega_{k}/(k_BT_0))-1)^{-1}$. $\omega_{k}$ is the magnon frequency, $k_B$ is the Boltzmann constant, and $T_0$ is the equilibrium temperature. When $f_{k} = f^{BE}_k(T_0)$, both the right and left hand sides of Eq.~(\ref{eq:mBTE}) vanish. 

For simplicity, we now normalize energy by $k_BT_0$ and then define 
\begin{equation}
n_{k}(t,\mathbf{x}) \equiv \Delta f_k(t,\mathbf{x})\text{sinh}(\hat{e}_k/2),
\end{equation}
where $\hat{e}_k =\hbar\omega_{k}/(k_BT_0)$ is the dimensionless magnon energy. This definition of the deviational distribution allows us to transform the magnon scattering term into a diagonalizable Hermitian matrix. Assuming $n_k(t,\mathbf{x})\text{sinh}^{-1}(\hat{e}_k/2)\ll f^{BE}_k(T_0)$, we can keep the terms only involving zeroth and first orders in $n_k(t,\mathbf{x})$ and, therefore, linearize the scattering operator around $f^{BE}_{k}(T_0)$. The linearized BTE in terms of  $n_{k}(t,\mathbf{x})$ can be written into the following form, 
\begin{equation}
\frac{\partial n_{k}}{\partial t}+\mathbf{v}_k\cdot \nabla n_{k} = -\frac{1}{\nu}\sum_{k'}\Omega_{kk'}n_{k'},
\label{eq:BTE_linearized}
\end{equation}
where $\nu$ is a normalized volume and $\Omega_{kk'}$ is the linear scattering operator acting on $n_{k'}$. This linearization of the scattering operator has been used in many studies of phonon transport and holds for small deviations from thermal equilibrium \cite{Ziman1960,chaput_direct_2013,cepellotti_thermal_2016,ward_ab_2009,broido_lattice_2005,li_shengbte:_2014}. The scattering matrix appearing in Eq.~(\ref{eq:BTE_linearized}) is in its most general form and describes all possible mechanisms by which a magnon excitation can be transferred from a state $k$ to a state $k'$ regardless of interaction mechanism. The matrix operator representing two-in/two-out magnon interactions is given in Appendix A. 

The matrix $\Omega$ has four key features: (1) it is real and symmetric, \emph{i.e.} $\Omega_{kk'} = \Omega_{k'k}$; (2) it is an even function of $k$, \emph{i.e.} $\Omega_{-k-k'} =\Omega_{kk'}$; (3) it is positive semi-definite, \emph{i.e.} $|\Omega_{kk'}|\geq 0$; and (4) due to the restriction from two-in/two-out scattering processes, it is summational invariant in both $u^0_k = \text{sinh}^{-1}(\hat{e}_k/2)$ and $u^1_k = \hat{e}_k\text{sinh}^{-1}(\hat{e}_k/2)$, \emph{i.e.}:
\begin{eqnarray}
&&\sum_{k'}\Omega_{kk'}u^0_{k'} = \sum_{k}u^0_k\Omega_{kk'} = \underline{0}, \label{eq:ParticleConv}\\
&&\sum_{k'}\Omega_{kk'}u^1_{k'} = \sum_{k}u^1_k\Omega_{kk'} = \underline{0}, \label{eq:Energyconv}
\end{eqnarray}
which represent conservation of total particle number and energy, correspondingly. The deviational particle number and nondimensional deviational energy are given by $\Delta N= \nu^{-1}\sum_k n_k(t,\mathbf{x})  u^0_k$ and $\Delta E  =\nu^{-1} \sum_k n_k(t,\mathbf{x})  u^1_k$, respectively. If $f_k$ is at a new thermodynamic equilibrium state described by a temperature $T$, the equilibrium deviational distribution is given as $n^{eq}_k(T) = u^1_k \Delta \hat{T}/4$, where $\Delta \hat{T} = |T-T_0|/T_0 \ll 1$.

The diagonal terms of $\Omega$ give the relaxation time, $\tau_k$, of each mode under the RTA, \emph{i.e.} $\partial n_{k}/\partial t|_{\mathrm{scattering}} = (n_k(t,\mathbf{x})-n^{eq}_k(T(t,\mathbf{x})))/\tau_k$. Under such an assumption, nonequilibrium magnon states decay back to equilibrium, independent of other magnons. However, it can be shown that the summational invariances of $u^0_k$ and $u^1_k$ cannot be simultaneously satisfied, meaning either the particle number or energy conservation has to be broken. This leads to an unphysical prediction of the nonequilibrium dynamics of magnons in Heisenberg antiferromagnets.

The goal of the following mathematical treatment is to treat the scattering matrix such that both energy and particle number are conserved and demonstrate that starting from a non-equilibrium distribution, for a closed system (no exchange with phonon bath or the environment through the boundaries), $n_k(t,\mathbf{x})$ will not return to an equilibrium distribution when conservation of both energy and particle number are present and the initial information of the distribution is preserved partially as $t \rightarrow \infty$. To show this, we start by solving Eq.~(\ref{eq:BTE_linearized}) using a spectral decomposition method \cite{guyer_solution_1966}. Due to the above mentioned properties of $\Omega$, we can deduce that there exists a complete set of eigenvectors such that 
\begin{equation}
\frac{1}{\nu}\sum_{k'}\Omega_{kk'}\theta^{\alpha}_{k'} = \frac{1}{\tau_\alpha}\theta^\alpha_k,
\end{equation}
where $ \tau_\alpha^{-1}$ is the eigenvalue (the lifetime of relaxons introduced by Cepellotti and Marzari \cite{cepellotti_thermal_2016})  and $\alpha$ ($\alpha = $ 0, 1, 2, 3 ...$N$, where $N$ is the dimension of the collision matrix) is the eigenvalue index of $\Omega$. The orthonormal condition and the scalar product are then defined as $\nu^{-1}\sum_{k}\theta^{\alpha}_{k}\theta^{\alpha'}_{k} = \langle \alpha|\alpha' \rangle = \delta_{\alpha\alpha'}$ and $\langle f|g \rangle=\nu^{-1}\sum_{k}f_{k}g_{k} = \langle g|f \rangle$. Then $n_{k}$ is expanded as 
\begin{equation}
n_{k}(t,\mathbf{x}) = \sum_{\alpha}g^{\alpha}(t,\mathbf{x})\theta^{\alpha}_{k},
\end{equation}
 where $g^{\alpha}(t,\mathbf{x}) = \langle n(t,\mathbf{x})|\theta^{\alpha} \rangle$ are unknown coefficients to be solved for.

Since $\Omega$ is real and symmetric, all of its eigenvectors must be real. Since $\Omega$ is an even function of $k$, its eigenvectors can be chosen to be either even or odd, \emph{i.e.} $\theta^\alpha_k = \pm \theta^\alpha_{-k}$. Because of its positive semi-definiteness, one can show that its eigenvalues are non-negative, \emph{i.e.} $\tau_\alpha \geq 0$ $\forall \alpha$, and there is a pair of degenerate eigenvectors with zero eigenvalue, which are labeled as $\alpha = 0, 1$. The associated eigenvectors are a linear superposition of $u^0_k$ and $u^1_k$, and therefore are even functions of $k$. 

It follows from Eq.~(\ref{eq:BTE_linearized}) and from the orthogonality and completeness of the eigenvectors that the coefficients $g^{\alpha}(t,\mathbf{x})$ are determined by the coupled set of equations,
\begin{eqnarray}\label{eq:BTE_coefficient}
\frac{\partial g^0}{\partial t} + \sum_{\beta, \mathrm{odd}} \langle 0|\mathbf{v}|\beta\rangle \cdot \nabla g^{\beta} &=& 0 \label{eq:BTE_zeroth} \\
\frac{\partial g^1}{\partial t} + \sum_{\beta,\mathrm{odd}} \langle 1|\mathbf{v}|\beta\rangle \cdot \nabla g^{\beta} &=& 0 \label{eq:BTE_first} \\
\frac{\partial g^{\alpha}}{\partial t}+\sum_{\beta} \langle \alpha|\mathbf{v}|\beta\rangle \cdot \nabla g^{\beta} &=& -\frac{g^{\alpha}}{\tau^{\alpha}},\ \alpha > 1 \label{eq:BTE_HO}
\end{eqnarray}
where the matrix elements of the group velocity are $\langle \alpha|\mathbf{v}|\beta\rangle = \nu^{-1}\sum_{k} \theta^\alpha_{k}\mathbf{v_{k}}\theta^\beta_{k}$. Since $\mathbf{v_{k}}$ is an odd function of $k$ ($\mathbf{v_{k}} = - \mathbf{v_{-k}}$), matrix elements connecting two eigenvectors with the same parity must be zero.

In general, solving the above system requires numerical discretization in time and space and matrix inversion. However, under the assumption of a closed boundary condition (no magnon flux in or out of the system), one is able to obtain an analytical solution for a volume integrated distribution function. Integrating Eqs.~(\ref{eq:BTE_zeroth})-(\ref{eq:BTE_HO}) over crystal volume, $V$, one obtain $\int_V \langle \alpha|\mathbf{v}|\beta \rangle \cdot \nabla g^{\beta} dV = \oint_S g^{\beta} \langle \alpha|\mathbf{v}|\beta\rangle \cdot \hat{\mathbf{n}} dS = 0$ for a closed boundary condition, where $\hat{\mathbf{n}}$ is the outward pointing unit normal at each point on the boundary $S$. Equations.~(\ref{eq:BTE_zeroth})-(\ref{eq:BTE_HO}) then are simplified into a set of decoupled first-order ordinary differential equations with solutions given as
\begin{eqnarray}
&&\tilde{g}_0(t) = \tilde{g}_0(t=0),\  \tilde{g}_1 = \tilde{g}_1(t=0),\nonumber \\
&&\tilde{g}_\alpha(t) = \tilde{g}_\alpha(t=0)\text{e}^{-t/\tau_\alpha} \ (\alpha>1), \label{eq:solution_g}
\end{eqnarray}
where the tilde means volume integrated. The time-dependent distribution function can be written as $\tilde{n}_k(t) = \tilde{g}_0(0)\theta^0_k+\tilde{g}_1(0)\theta^1_k+\sum_{\alpha>1}\tilde{g}_\alpha(0)\text{e}^{-t/\tau_\alpha}\theta^\alpha_k$. 
Therefore, the stationary state solution is given by
\begin{equation}
\tilde{n}^{ss}_k = \tilde{n}_k(t\rightarrow \infty) = \tilde{g}_0(0)\theta^0_k+\tilde{g}_1(0)\theta^1_k. \label{eq:ss_solution}
\end{equation}
To gain more physical insights into the stationary state solution, we write $\theta^0_k$ as 
\begin{equation}
\theta^0_k=au^0_k+bu^1_k.
\end{equation}
Due to orthonormality between the eigenvectors, $\theta^1_k$ can be constructed as $\theta^1_k =\chi^{-1/2} [u^0_k-\langle u^0|\theta^0\rangle \theta^0_k]$ with the normalization factor, $\chi$. For simplicity, $\theta^1_k$ can be written as 
\begin{equation}
\theta^1_k = cu^0_k-du^1_k,
\end{equation}
where 
\begin{equation}
c = \frac{a \langle u^0| u^1\rangle+b \langle u^1|u^1\rangle}{\sqrt{\langle u^0|u^0\rangle \langle u^0|u^0\rangle- \langle u^0| u^1\rangle^2}}, 
\end{equation}
and
\begin{equation}
d = \frac{b \langle u^0| u^1\rangle+a \langle u^0|u^0\rangle}{\sqrt{\langle u^0|u^0\rangle \langle u^0|u^0\rangle- \langle u^0| u^1\rangle^2}}.
\end{equation} 

Assuming a volumetric generation of magnons at $t=0$ with an initial distribution $\tilde{n}^i_k$, the total number of generated particles is given as $\Delta N = \langle \tilde{n}^i| u^0 \rangle$ and the total dimensionless energy generated as $\Delta E = \langle \tilde{n}^i | u^1\rangle$. Then, we have $\tilde{g}_0(0) = a \Delta N + b\Delta E$ and $\tilde{g}_1(0) = c\Delta N - d\Delta E$. Then the stationary state solution can be rewritten as 
\begin{eqnarray}
\tilde{n}^{ss}_k &=& (a^2 \Delta N+ab\Delta E +c^2\Delta N-cd\Delta E) u^0_k \nonumber \\ 
&+& (ab\Delta N + b^2 \Delta E-cd\Delta N+d^2 \Delta E) u^1_k.
\label{eq:n_ss}
\end{eqnarray}

Using the identities given in Appendix B, we have $\langle \tilde{n}^{ss}|u^0\rangle = \Delta N$ and $\langle \tilde{n}^{ss}|u^1\rangle = \Delta E$, indicating that both particle number and energy are conserved. The only way for $\tilde{n}_k^{ss}$ to reach an equilibrium distribution is to set $a^2 \Delta N+ab\Delta E +c^2\Delta N-cd\Delta E = 0$, leading to $\Delta N/\Delta E = \langle u^0|u^1\rangle/\langle u^1|u^1\rangle$ and $\tilde{n}^i_k \propto u^1_k$. This shows that for a closed magnon system with magnon-magnon interactions restricted with conservation of both energy and particle number, unless the system starts with an equilibrium initial distribution, the magnon population will always remain out-of-equilibrium. 

If the particle number is not conserved, $u^0_k$ is no longer a summational invariant of the collision matrix, $\Omega$, and the time-dependent distribution function becomes $\tilde{n}_k(t) =\tilde{g}_1(0)u^1_k+\sum_{\alpha>1}\tilde{g}_\alpha(0)\text{e}^{-t/\tau_\alpha}\theta^\alpha_k$. Then, $\tilde{n}^{ss}_k$  will always be proportional to the equilibrium distribution, $u^1_k$, meaning the system will reach equilibrium as $t \rightarrow \infty$, similar to that for a typical phonon system mediated by three-phonon interactions.

\section{Dynamics of nonequilibrium magnons}

\begin{figure}
\includegraphics[scale = 0.45]{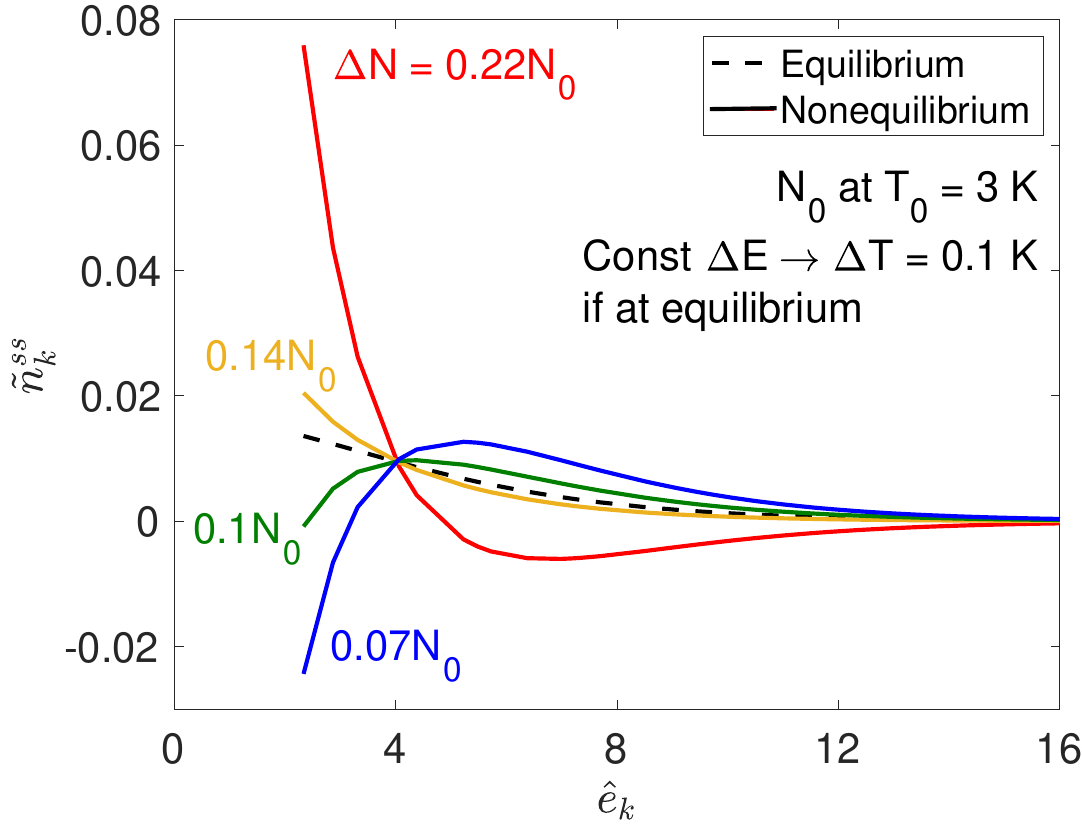}
\caption{The stationary state distributions, $\tilde{n}^{ss}_{k}$, as a function of dimensionless magnon energy, $\hat{e}_k$, for four initial conditions (solid curves). For all four initial distributions, the total $\Delta E$ is kept as a constant, around 15\% of the total energy at 3 K, $E_0 = \nu^{-1}\sum_k\hbar\omega_kf^{BE}_k(T_0=3K)$. Depending on how many states are initially excited, $\Delta N$ varies for the four cases: $\Delta N = 0.22 N_0$ (red), $0.14 N_0$ (yellow), $0.09 N_0$ (green), and $0.07 N_0$ (blue), where $N_0 = \nu^{-1}\sum_k f^{BE}_k(T_0 =3K)$ is the total number of particles at 3 K. The dashed curve gives the equilibrium deviational distribution at a given $\Delta E$, which corresponds to a temperature rise of $\Delta T$ = 0.1 K.}
\label{fig:Distribution}
\end{figure}

We now apply the above calculation to Rb$_2$MnF$_4$, a quasi-2D square lattice S = 5/2 Heisenberg antiferromagnet. The Hamiltonian of this material is well understood \cite{huberman_two-magnon_2005}. Its magnon dispersion and damping mechanism have been extensively studied both theoretically and experimentally\cite{huberman_study_2008,bayrakci_lifetimes_2013,halperin_hydrodynamic_1969,harris_dynamics_1971}. Magnon transport is confined to the $ab$ plane and ion anisotropy leads to a magnon energy gap around 0.6 meV at 3 K. At $T \ll T_N = 38$ K, the linewidth broadening of one-magnon scattering intensity by neutrons has been confirmed to be dominated by two-in/two-out magnon-magnon interactions and phonon-magnon interactions are negligible \cite{bayrakci_lifetimes_2013, huberman_study_2008}. The following calculations are performed at $T_0 = 3$ K and all the parameters necessary to evaluate the dispersion and scattering matrix are given in Refs.~\cite{huberman_two-magnon_2005, bayrakci_lifetimes_2013}.

We calculate the matrix elements of $\Omega$ and numerically determine the values of $a$ and $b$ in $\theta^0_k$ by finding the eigenvalues and eigenvectors of $\Omega$. For a given initial distribution, $\tilde{n}_i$, we are now able to evaluate its corresponding stationary state solution using Eq.~(\ref{eq:n_ss}). Figure.~\ref{fig:Distribution} gives the stationary state distribution functions, $\tilde{n}^{ss}_{k}$, for four initial conditions (solid curves). For all four initial distributions, the total $\Delta E$ is kept constant. Depending on how many states are initially excited, $\Delta N$ varies for the four cases. The dashed curve in Fig.~\ref{fig:Distribution} gives the equilibrium deviational distribution, $\tilde{n}^{eq}_k(T)$, at the given $\Delta E$, which corresponds to a temperature rise of 0.1 K. The crossing of all the distributions in Fig.~\ref{fig:Distribution} occurs at $\hat{e}_k = \langle u^1|u^1\rangle/\langle u^0| u^1\rangle$ where all the terms involving $\Delta N$ in Eq.~(\ref{eq:n_ss}) cancel out. From Eq.~(\ref{eq:n_ss}) and Fig.~\ref{fig:Distribution}, we have shown that the final stationary state solutions depend on the initial conditions through the values of $\Delta N$ and $\Delta E$ but it cannot distinguish two initial distributions that give the same $\Delta N$ and $\Delta E$. Therefore, the stationary state solution contains partial information about the initial states.

\begin{figure*}
\includegraphics[scale = 0.45]{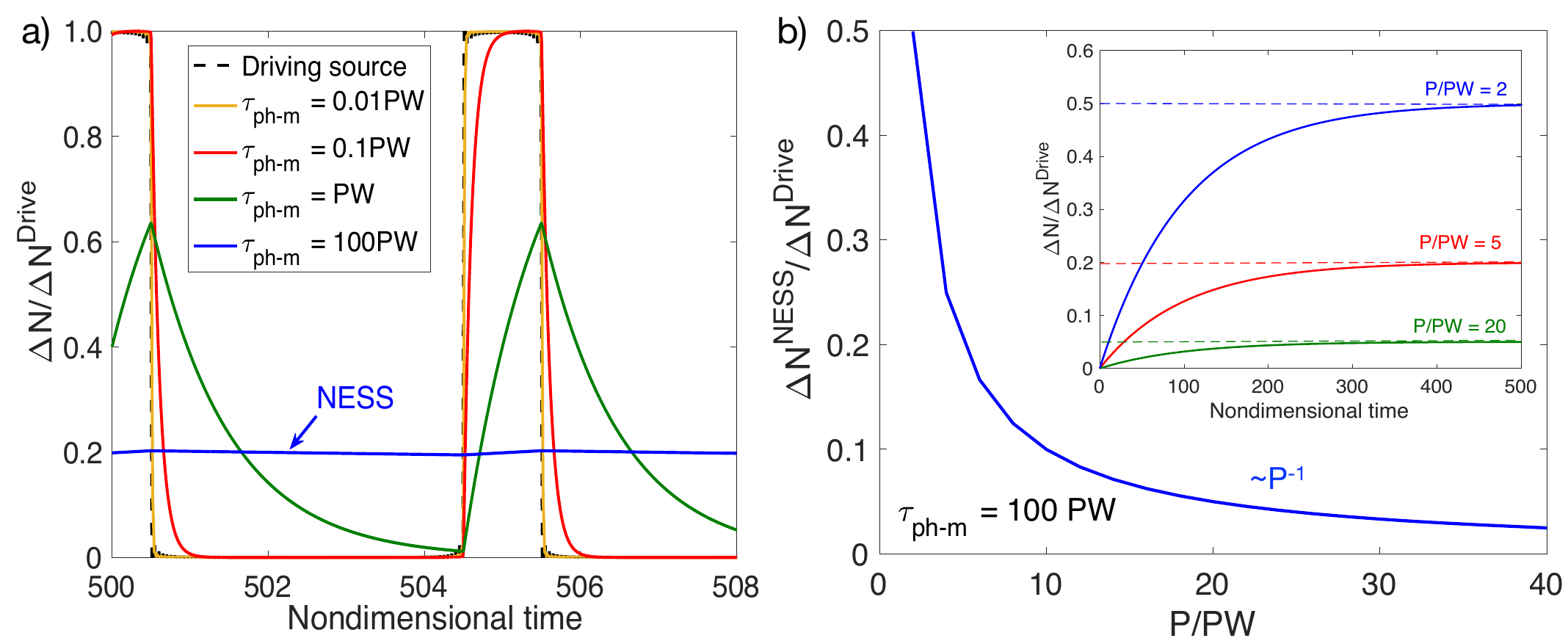}
\caption{ (a) $\Delta N(t)$ (solid curves) under a square-wave driving source, $\Delta N^{\text{Drive}}(t)$ (dashed curve). Time is nondimensionalized by a fixed pulse width (PW) and the period, P, of the source is fixed at P = 5 PW. The dynamical behavior of $\Delta N(t)$ changes with phonon-magnon coupling time ($\tau_{ph-m}$): $\tau_{ph-m} = 0.1$ PW (yellow),  PW (red),  10 PW (green), 100 PW (blue). When $\tau_{ph-m} \gg$ PW (blue line), the magnon system reaches a nonequilibrium steady state and nonequilibrium magnon population will be maintained as long as the driving source lasts. (b) Nonequilibrium steady state value, $\Delta N^{\text{NESS}}$ is inversely proportional to the period, P. Inset: $\Delta N(t)$ under a square-wave driving source at three different periods. }
\label{fig:NESS}
\end{figure*}

To analyze the driven-dissipative behavior of the dynamics, we now add a phenomenological magnon-phonon coupling as a prototypical non-conserving interaction and external driving terms to Eq.~(\ref{eq:BTE_linearized}) and the volume-integrated BTE then becomes 
\begin{equation}
\frac{\partial \tilde{n}_{k}}{\partial t} = -\frac{1}{\nu}\sum_{k'}\Omega_{kk'}\tilde{n}_{k'}-\frac{\tilde{n}_k-\tilde{n}^{eq}_k(T)}{\tau_{ph-m}}+F_k(t),
\label{eq:BTE_n}
\end{equation}
where $\tau_{ph-m}$ is the relaxation time of magnons due to phonon-magnon interaction and $F_k(t)$ is an external driving source. To ease the derivation but still maintain the physical picture, $\tau_{ph-m}$ is assumed a constant for all magnon modes. After application of the eigendecomposition method as described above, the governing equations for $\tilde{g}^\alpha (\alpha \geq 2)$ become 
\begin{equation}
\frac{\partial\tilde{g}^\alpha}{\partial t} = -\frac{\tilde{g}^\alpha}{\tau_\alpha}-\frac{\tilde{g}^{\alpha}}{\tau_{ph-m}}+F^\alpha(t),
\end{equation}
where $F^\alpha(t) = \langle F(t)|\theta^\alpha\rangle$. The solution to this first-order differential equation is then given by  $\tilde{g}^\alpha(t) = \int^t_0 F^\alpha(t-t') e^{-t'/\tau_\alpha-t'/\tau_{ph-m}}dt'$. From the numerical evaluation, the largest value of $\tau_\alpha$ for Rb$_2$MnF$_4$ is around $\sim 1~\mu$s. At T $\ll$ T$_N$, the relaxation time due to phonon-magnon interaction in Rb$_2$MnF$_4$ is assumed much longer than 1~$\mu$s. Then the decay dynamics of $\tilde{g}^\alpha(t)$ is determined by $\tau_\alpha$. The dynamics of $\Delta N$ and $\Delta E$, on the other hand, is determined by $\tau_{ph-m}$, given by the following equations
\begin{eqnarray}
a\left(\frac{\partial \Delta N}{\partial t} +\frac{\Delta N}{\tau_{ph-m}}\right) &+& b\left(\frac{\partial \Delta E}{\partial t}+\frac{\Delta E }{\tau_{ph-m}}\right) \nonumber \\
&=& \frac{q\Delta\hat{T}}{\tau_{ph-m}}+F^0(t), \label{eq:g0_driven}\\
c\left(\frac{\partial \Delta N}{\partial t} +\frac{\Delta N}{\tau_{ph-m}}\right) &+& d\left(\frac{\partial \Delta E}{\partial t}+\frac{\Delta E }{\tau_{ph-m}}\right) \nonumber \\
& =& \frac{p\Delta \hat{T}}{\tau_{ph-m}}+F^1(t), \label{eq:g1_driven}
\end{eqnarray}
where $F^{0,1}(t) = \langle F(t)|\theta^{0,1}\rangle$, $q = a/[4(ad-bc)]$ and $p = c/[4(bc-ad)]$. If $F^\alpha(t)$ is a driving source that only affects magnons and $\Delta \hat{T}$ is a time-independent quantity determined by phonons, then Eqs.~(\ref{eq:g0_driven}) and (\ref{eq:g1_driven}) can be decoupled into two first-order differential equations and their solutions have a general form of $\int^t_0 \Delta N^{\text{Drive}}(t-t')e^{-t'/\tau_{ph-m}}dt'+constant$, where $\Delta N^{\text{Drive}}(t)$ is an external driving function in a general form. 

Figure~\ref{fig:NESS} shows the dynamics of $\Delta N(t)$ under a square-wave driving source, $\Delta N^{\text{Drive}}(t) = \Delta N^{\text{Drive}}(t+P)$, where P is the source period. In this study, we fix the pulse width (PW) of the square-wave source. The ratio of $\tau_{ph-m}$ and PW determines the time-dependent behavior of $\Delta N(t)$. When $\tau_{ph-m} \ll$ PW (yellow curve in Fig.~\ref{fig:NESS}(a)), due to a fast magnon-phonon coupling time relative to the active pumping time, the magnon system quickly relaxes back to equilibrium once the driving source disappears. When $\tau_{ph-m} \sim$ PW (red and green curves Fig.~\ref{fig:NESS}(a)), decay of the nonequilibrium magnon population is observed between two driving events and the decay time is determined by $\tau_{ph-m}$. Observation of this dynamical behavior can be used to determine the coupling time of magnon-phonon interactions experimentally.  When $\tau_{ph-m} \gg$ PW (blue curve in Fig.~\ref{fig:NESS}(a)), the magnon system reaches a nonequilibrium steady state and a nonequilibrium magnon population will be maintained as long as the driving source lasts. Once a nonequilibrium steady state condition is reached ($\tau_{ph-m} \gg$ PW), changing the source period (P) only affects the nonequilibrium steady state value as shown in the inset of Fig.~\ref{fig:NESS}(b). Fig.~\ref{fig:NESS}(b) shows the nonequilibrium steady state value is inversely proportional to P. As $P \rightarrow \infty$, $\Delta N^{\mathrm{NESS}}$ approaches zero and the system is at equilibrium.

Candidate experimental methods that may be suitable for observing the magnon NESS include but are not limited to Raman\cite{Yang_UltrafastMangetic_2020}, Brillouin light\cite{demokritov_boseeinstein_2006, bozhko_supercurrent_2016, borisenko_direct_2020} and inelastic neutron scatterings\cite{hua2023implementation}. All three methods mentioned here have been routinely used to measure magnon properties at thermodynamic equilibrium and are feasible to integrate a periodic pumping source, \emph{i.e.}, a laser pump. While optical spectroscopies probe nonequilibrium modes at zero momentum, inelastic neutron scattering allows access to the momentum-energy space of nonequilibrium magnons in a quantum magnet.   According to the theory developed in this work, by varying the pumping duration of the laser pulses, different decay behaviors of nonequilibrium magnons, similar to that shown in Fig.~\ref{fig:NESS}, should be observed in these pump-probe experiments.

\section{conclusion}

We studied the dynamics of nonequilibrium magnons in a gapped Heisenberg antiferromagnet in the framework of Boltzmann transport theory. When solving the Boltzmann transport equation, the macroscopic conservation laws mandated by the elastic pairwise collisions of magnons in a Heisenberg model were strictly obeyed. We found that intrinsic magnon-magnon interactions are not sufficient to re-equilibrate nonequilibrium states unless the magnon system undergoes some non-conserving interactions, \emph{i.e.} coupled to a thermal bath of phonons. Therefore, the necessary condition to achieve the NESS of magnons is that any non-conserving interaction has a much slower relaxation time than the intrinsic magnon-magnon interaction time.  The sufficient condition to observe the magnon NESS in such a magnet is to pump nonequilibrium magnons to the system periodically, \emph{i.e.} via laser pumping, at a rate faster than the relaxation times of non-conserving interactions. This sufficient condition provides a route to experimentally determine the coupling time of magnon-phonon interactions, a quantity that remains unknown in most magnets.

As a final remark, there is a class of quantum magnets like gapped Heisenberg antiferromagnets \cite{harris_dynamics_1971} that satisfy the two-in/two-out magnon-magnon scattering rule. In these material systems, nonequilibrium magnon populations, once excited, will not relax back to thermodynamic equilibrium unless there are other energy leakage channel, \emph{i.e}, coupling with a phonon bath. These material systems provide a natural platform to study thermalization processes, nonequilibrium statistical physics, and transport mediated by the nonequilibrium states, all of which are important to both fundamental science and technological applications.

\section*{Acknowledgments}
This work was supported by the U. S. Department of Energy, Office of Science, Basic Energy Sciences, Materials Sciences and Engineering Division and National Quantum Information Science Research Centers, Quantum Science Center. 

\begin{appendices}
\begin{widetext}
\numberwithin{equation}{section}
\section{Linearized scattering matrix}\label{AppendixI}

For two-in/two-out magnon-magnon scattering, following the expression given by Harris \emph{et. al.} \cite{harris_dynamics_1971}, the collision operator of the Boltzmann transport equation is written as
\begin{eqnarray}
\frac{\partial f_k}{\partial t}\biggm|_{scattering} &=&\sum_{p,s}\frac{\pi H_E^2}{16 S^2 N^2}M_{22}(k,p,r,s)\delta(\omega_k+\omega_p-\omega_r-\omega_s)\Delta(\boldsymbol{q}_k+\boldsymbol{q}_p-\boldsymbol{q}_r-\boldsymbol{q}_s)  \nonumber \\
&\times&  \left[(f_k+1)(f_p+1)f_r f_s - f_k f_p (f_r+1)(f_s+1)\right],
\label{eq:scattering}
\end{eqnarray}
where $M_{22}(k,p,r,s)$ in Eq.~(\ref{eq:scattering}) is referred to as the "matrix element" by Harris \emph{et. al.} and its derivation can be found in Ref.~\cite{bayrakci_lifetimes_2013}. $H_E = 2JzS$,  where $J$ is the spin exchange strength, S is the spin moment, and $z$ is a quantum renormalization factor. The Kronecker delta $\Delta$ is zero unless its argument is zero or a reciprocal lattice vector, in which case it takes the value $1$.

By defining $f_{k} \equiv  f^{BE}_{k}(T_0)+n_k\text{sinh}^{-1}(\hat{e}_k/2)$ and assuming $n_k\text{sinh}^{-1}(\hat{e}_k/2)\ll f^{BE}_k(T_0)$, Eq.~\ref{eq:scattering} is linearized into
\begin{eqnarray}\nonumber
&&\frac{\partial n_{k}}{\partial t}\biggm|_{s} = \sum_{p,s}\frac{\pi H_E^2}{16 S^2 N^2}M_{22}(k,p,r,s)\delta(\omega_k+\omega_p-\omega_r-\omega_s)\Delta(\boldsymbol{q}_k+\boldsymbol{q}_p-\boldsymbol{q}_r-\boldsymbol{q}_s) \\
&& \times \left[\frac{n_k\text{sinh}\left(\frac{\hat{e}_k}{2}\right)}{4\text{sinh}\left(\frac{\hat{e}_p}{2}\right)\text{sinh}\left(\frac{\hat{e}_r}{2}\right)\text{sinh}\left(\frac{\hat{e}_s}{2}\right)}+\frac{n_p}{4\text{sinh}\left(\frac{\hat{e}_r}{2}\right)\text{sinh}\left(\frac{\hat{e}_s}{2}\right)}-\frac{n_r}{4\text{sinh}\left(\frac{\hat{e}_p}{2}\right)\text{sinh}\left(\frac{\hat{e}_s}{2}\right)}-\frac{n_s}{4\text{sinh}\left(\frac{\hat{e}_p}{2}\right)\text{sinh}\left(\frac{\hat{e}_r}{2}\right)}\right] \label{eq:LinearizedFourMagnons}
\end{eqnarray}
Using the fact that for each collision process ($k+p\rightarrow r+s$), we can find three other decay processes ($p+k\rightarrow r+s$, $r+s\rightarrow k+p$, and $s+r\rightarrow k+p$). The matrix form of this linearized scattering operator is therefore symmetric, written as
\begin{equation}
\frac{\partial n_k}{\partial t}\biggm|_{s} = \frac{1}{\nu}\sum_{k'}\Omega_{kk'}n_{k'}
\end{equation}
where
\begin{eqnarray}
\Omega_{kk} &=& \sum_{p,s}\frac{\pi H_E^2}{16 S^2 N^2}M_{22}(k,p,r,s)\delta(\omega_k+\omega_p-\omega_r-\omega_s)\Delta(\boldsymbol{q}_k+\boldsymbol{q}_p-\boldsymbol{q}_r-\boldsymbol{q}_s)\frac{\text{sinh}\left(\frac{\hat{e}_k}{2}\right)}{4\text{sinh}\left(\frac{\hat{e}_p}{2}\right)\text{sinh}\left(\frac{\hat{e}_r}{2}\right)\text{sinh}\left(\frac{\hat{e}_s}{2}\right)}\\
\Omega_{pp} &=& \sum_{r,s}\frac{\pi H_E^2}{16 S^2 N^2}M_{22}(k,p,r,s)\delta(\omega_k+\omega_p-\omega_r-\omega_s)\Delta(\boldsymbol{q}_k+\boldsymbol{q}_p-\boldsymbol{q}_r-\boldsymbol{q}_s)\frac{\text{sinh}\left(\frac{\hat{e}_p}{2}\right)}{4\text{sinh}\left(\frac{\hat{e}_k}{2}\right)\text{sinh}\left(\frac{\hat{e}_r}{2}\right)\text{sinh}\left(\frac{\hat{e}_s}{2}\right)}\\
\Omega_{rr} &=& \sum_{p,s}\frac{\pi H_E^2}{16 S^2 N^2}M_{22}(k,p,r,s)\delta(\omega_k+\omega_p-\omega_r-\omega_s)\Delta(\boldsymbol{q}_k+\boldsymbol{q}_p-\boldsymbol{q}_r-\boldsymbol{q}_s)\frac{\text{sinh}\left(\frac{\hat{e}_r}{2}\right)}{4\text{sinh}\left(\frac{\hat{e}_k}{2}\right)\text{sinh}\left(\frac{\hat{e}_p}{2}\right)\text{sinh}\left(\frac{\hat{e}_s}{2}\right)}\\
\Omega_{ss} &=&\sum_{p,r}\frac{\pi H_E^2}{16 S^2 N^2}M_{22}(k,p,r,s)\delta(\omega_k+\omega_p-\omega_r-\omega_s)\Delta(\boldsymbol{q}_k+\boldsymbol{q}_p-\boldsymbol{q}_r-\boldsymbol{q}_s)\frac{\text{sinh}\left(\frac{\hat{e}_s}{2}\right)}{4\text{sinh}\left(\frac{\hat{e}_k}{2}\right)\text{sinh}\left(\frac{\hat{e}_p}{2}\right)\text{sinh}\left(\frac{\hat{e}_r}{2}\right)}\\
\Omega_{kp} &=& \Omega_{pk} = \sum_{r}\frac{\pi H_E^2}{16 S^2 N^2}M_{22}(k,p,r,s)\delta(\omega_k+\omega_p-\omega_r-\omega_s)\Delta(\boldsymbol{q}_k+\boldsymbol{q}_p-\boldsymbol{q}_r-\boldsymbol{q}_s)\frac{1}{4\text{sinh}\left(\frac{\hat{e}_r}{2}\right)\text{sinh}\left(\frac{\hat{e}_s}{2}\right)}\\
\Omega_{ks} &=& \Omega_{sk} = -\sum_{p}\frac{\pi H_E^2}{16 S^2 N^2}M_{22}(k,p,r,s)\delta(\omega_k+\omega_p-\omega_r-\omega_s)\Delta(\boldsymbol{q}_k+\boldsymbol{q}_p-\boldsymbol{q}_r-\boldsymbol{q}_s)\frac{1}{4\text{sinh}\left(\frac{\hat{e}_p}{2}\right)\text{sinh}\left(\frac{\hat{e}_r}{2}\right)}\\
\Omega_{kr} &=& \Omega_{rk} = -\sum_{p}\frac{\pi H_E^2}{16 S^2 N^2}M_{22}(k,p,r,s)\delta(\omega_k+\omega_p-\omega_r-\omega_s)\Delta(\boldsymbol{q}_k+\boldsymbol{q}_p-\boldsymbol{q}_r-\boldsymbol{q}_s)\frac{1}{4\text{sinh}\left(\frac{\hat{e}_p}{2}\right)\text{sinh}\left(\frac{\hat{e}_s}{2}\right)}\\
\Omega_{pr} &=& \Omega_{rp} = -\sum_{s}\frac{\pi H_E^2}{16 S^2 N^2}M_{22}(k,p,r,s)\delta(\omega_k+\omega_p-\omega_r-\omega_s)\Delta(\boldsymbol{q}_k+\boldsymbol{q}_p-\boldsymbol{q}_r-\boldsymbol{q}_s)\frac{1}{4\text{sinh}\left(\frac{\hat{e}_k}{2}\right)\text{sinh}\left(\frac{\hat{e}_s}{2}\right)}\\
\Omega_{ps} &=& \Omega_{sp} = -\sum_{r}\frac{\pi H_E^2}{16 S^2 N^2}M_{22}(k,p,r,s)\delta(\omega_k+\omega_p-\omega_r-\omega_s)\Delta(\boldsymbol{q}_k+\boldsymbol{q}_p-\boldsymbol{q}_r-\boldsymbol{q}_s)\frac{1}{4\text{sinh}\left(\frac{\hat{e}_k}{2}\right)\text{sinh}\left(\frac{\hat{e}_r}{2}\right)}\\
\Omega_{rs} &=& \Omega_{sr} = \sum_{p}\frac{\pi H_E^2}{16 S^2 N^2}M_{22}(k,p,r,s)\delta(\omega_k+\omega_p-\omega_r-\omega_s)\Delta(\boldsymbol{q}_k+\boldsymbol{q}_p-\boldsymbol{q}_r-\boldsymbol{q}_s)\frac{1}{4\text{sinh}\left(\frac{\hat{e}_k}{2}\right)\text{sinh}\left(\frac{\hat{e}_p}{2}\right)}
\end{eqnarray}

\section{Eigenvector Identities}\label{AppendixII}
Due to the orthonormal relation of $\theta^0_k$ and $\theta^1_k$ (Eqns. 14 and 15), we can derive the following identities of the coefficients (a,b,c,d) in $\theta^0_k$ and $\theta^1_k$:
\begin{eqnarray}
cd - ab &=& \frac{\langle u^0 | u^1\rangle}{\langle u^0|u^0\rangle \langle u^1|u^1\rangle-\langle u^0| u^1\rangle^2} \geq 0,\\
a^2+c^2 &=& \frac{ \langle u^1|u^1\rangle}{\langle u^0|u^0\rangle \langle u^1|u^1\rangle-\langle u^0| u^1\rangle^2} \geq 0,\\
b^2+d^2 &=& \frac{ \langle u^0|u^0\rangle}{\langle u^0|u^0\rangle \langle u^1|u^1\rangle-\langle u^0| u^1\rangle^2} \geq 0.
\end{eqnarray}
Using the above identities and Eq.~(\ref{eq:n_ss}), the total deviational particle number and energy at the stationary state are then given as 
\begin{eqnarray}
\langle \tilde{n}^{ss}|u^0 \rangle = [(a^2+c^2)\langle u^0|u^0\rangle+(ab-cd)\langle u^0 | u^1\rangle]\Delta N + [(ab-cd)\langle u^0|u^0\rangle + (b^2+d^2)\langle u^0 | u^1\rangle]\Delta E  = \Delta N, \\
\langle \tilde{n}^{ss}|u^1 \rangle = [(a^2+c^2)\langle u^0 | u^1\rangle+(ab-cd)\langle u^1|u^1\rangle]\Delta N + [(ab-cd)\langle u^0 | u^1\rangle + (b^2+d^2)\langle u^1|u^1\rangle]\Delta E  = \Delta E.
\end{eqnarray}

\end{widetext}
\end{appendices}


%

\end{document}